\documentclass[a4paper,10pt,twoside]{cpc-hepnp}

\usepackage{multicol}
\usepackage{graphicx}
\usepackage{booktabs}
\usepackage{amssymb,bm,mathrsfs,bbm,amscd}
\usepackage[tbtags]{amsmath}
\usepackage{lastpage}
\usepackage{CJK}

\begin{document}
\begin{CJK*}{GBK}{song}

\fancyhead[c]{\small Chinese Physics C~~~Vol. 42, No. 1 (2018)
010201} \fancyfoot[C]{\small 010201-\thepage}

\footnotetext[0]{Received 23 August 2018}
\title{
Simple description of the temperature dependencies of the width of fission-fragment mass yield in $^{197}$Au and $^{209}$Bi at intermediate energies
}%

\author{
V. Yu. Denisov$^{1,2,3;1)}$\email{denisov@kinr.kiev.ua}%
\quad O. A. Belyanovska$^{2}$
\quad V. P. Khomenkov$^{1,2,4}$,
\quad I. Yu. Sedykh$^{5}$,
\quad K. M. Sukhyy$^{2}$
}%

\maketitle

\address{%
$^{1}$ Institute for Nuclear Research, Prospect Nauki 47,
03028 Kiev, Ukraine \\
$^{2}$ Ukrainian State University of Chemical Technology, Prospect Gagarina 8,
49005 Dnepr (Dnepropetrovsk), Ukraine\\
$^{3}$ Faculty of Physics, Taras Shevchenko National University of Kiev, Prospect Glushkova 2, 03022 Kiev, Ukraine\\
$^{4}$ National Technical University of Ukraine "Igor Sikorsky Kyiv Polytechnic Institute", Prospect Peremogy 37, 03056 Kiev, Ukraine\\
$^{5}$ Financial University, Leningradsky Prospect 49, 125993 Moscow, Russia
}%

\begin{abstract}

Simple approach is used to a description of the experimental data for the widths of fission fragment mass yields in $^{197}$Au and $^{209}$Bi at low and intermediate energies. This approach based on both the expression for the temperature dependence of the width of fission fragment mass yield and the mass of the most probable fragment. The expression for the width of fission fragment mass yield depends on the mass of the most probable fragment, the surface terms of the energy level density parameter, the temperature and the stiffness parameter of the potential related to mass-asymmetric degree of freedom. It is shown that the contribution of the surface term of the energy level density parameter into the width is important for description of the experimental data in wide range of energies.

\end{abstract}

\begin{keyword}
fission fragment mass yield; width of fission fragment mass yield; temperature dependence; fission
\end{keyword}

\begin{pacs}
24.75.+i;	25.85.+k	
\end{pacs}

\footnotetext[0]{\hspace*{-3mm}\raisebox{0.3ex}{$\scriptstyle\copyright$}2018
Chinese Physical Society and the Institute of High Energy Physics
of the Chinese Academy of Sciences and the Institute
of Modern Physics of the Chinese Academy of Sciences and IOP Publishing Ltd}%

\begin{multicols}{2}

\section{Introduction}

The mass yield of fission fragments are most studied feature of the nuclear fission \cite{hw,f,nf,s1964,nix,vh,nr,andr,iors,agpps,ap,itkis90,expAu,expBi1,expBi2,expBi3,expBi4,expBi5,expBi6,o,nnn,wahl,liu,wang,spy,js,mr,t,i,w,m,sjas,dms,ds,dbkss}. The mass yield is often described by the several Gaussians \cite{vh,iors,wahl,liu}, which have the corresponding widths $\sigma^2$. The width of fission-fragment mass yield is a very important quantity of the fission process \cite{nf,s1964,nix,vh,expAu,expBi1,expBi2,expBi3,expBi4,expBi5,expBi6,o,iors,agpps,ap,t,m,liu,wang,js,sjas,dbkss}.

Neuzil and Fairhall proposed oversimplified an empirical relationship for the width of fission-fragment mass yield \cite{nf}. Strutinsky in the framework of statistical model found that the width of fission-fragment mass yield is proportional to the temperature $T$ of the fissioning system at the saddle point, the square of number nucleons in nuclei $A^2$ and inverse-proportional to the stiffness parameter $C$ of the potential related to mass-asymmetric degree of freedom \cite{s1964}. Nix shown that this width in very low-energy fission is strongly influenced by the zero-point motion of the corresponding quantum oscillators \cite{nix}. The zero-point motion of the corresponding quantum oscillators is important only in the very low temperature limit, where the values of the width evaluated using Nix expression are larger than the ones obtained by Strutinsky formula. However, Strutinsky and Nix results are the same for a high temperatures. The numerical studies of the width in the framework of the diffusion model of fission have been also done in Refs. \cite{agpps,ap,m}.

The mass distribution of the fission fragment of highly-excited fissioning nuclei with the number of nucleons $A\lesssim 220$ is related to the two-body saddle point \cite{dms}. Recently, we have found new simple expression for the width of fission-fragment mass yield at moderate and high excitation energies of the fissioning system in the framework of the statistical approach \cite{dbkss}. In contrast to previous results \cite{s1964,nix,iors} we take into account both the volume and surface terms of the energy level density parameter \cite{ripl3} in our statistical approach for the width. As a result, the temperature dependence of the width is changed due to the contribution of the surface term of the energy level density parameter. However, our expression for the width is transformed into the Strutinsky result in low temperature limit. Note that we consider temperatures at which the zero-point motion of the corresponding quantum oscillators is negligible. The difference between our and Strutinsky results rises with temperature of fissioning nucleus and becomes important at temperatures $T \gtrsim 1$ MeV. The width of fission-fragment mass yield evaluated by using the volume and surface terms of the energy level density parameter is smaller than the one obtained at the same values of parameters and using the volume term only  \cite{dbkss}.

In this paper, we discuss the energy dependence of the width of fission-fragment mass yield in the photo-fission of $^{197}$Au and $^{209}$Bi in the wide energy interval. The experimental energy dependence of the width was measured in the photo-fission of $^{197}$Au target by the bremsstrahlung of end-point energies 300--1100 MeV in Ref. \cite{expAu}. The data for the widths obtained in the photo-fission of $^{209}$Bi target by the bremsstrahlung of end-point energies 40--1100 MeV are given in Refs. \cite{expBi1,expBi2,expBi3,expBi4,expBi5}. These data have not analyzed using expressions for the width from Refs. \cite{s1964,dbkss} up to now. We also take into account the effect of neutron evaporation on the width of fission-fragment mass yield and apply expressions for the width from Refs. \cite{s1964,dbkss}.

The paper is organized as follows. Two expressions for the fission width \cite{s1964,dbkss} are shortly described in Sec. 2. A discussion of results and conclusions are given in Sec. 3.

\section{The expressions for the width of fission-fragment mass yield}

As we pointed in the introduction, the mass yields of fission fragments are often described by the several Gaussians \cite{vh,iors,wahl,liu}. The width $\sigma^2$ described the yield of fission fragment with mass $A_1$ at the symmetric fission of nucleus with $A$ nucleons is presented in the Gaussian $\exp{ \left[ - (A_1-A/2)^2/\sigma^2 \right]}$ \cite{vh,wahl,dbkss}.

The width of mass distribution of fission fragments obtained in Ref. \cite{dbkss} is given by
\begin{eqnarray}
\sigma^2 = \frac{2 A^2 T}{C + 2 \kappa A^{2/3} T^2}.
\end{eqnarray}
Here $A$ is the number of nucleons of the fissioning system in the two-body saddle point, $T$ is the temperature of the system, which is related to the excitation energy $E^\star=a_{\rm s}(A) T^2$ of the system of two identical fission fragments at the saddle point, $C$ is the stiffness parameter of the potential related to the mass-asymmetric degree of freedom in the saddle point,
\begin{eqnarray}
\kappa = \frac{4 \cdot 2^{1/3}}{9} \beta, \\
a_{\rm sp}(A) = \alpha A + 2^{1/3} \beta A^{2/3},
\end{eqnarray}
is the asymptotic value of the energy level density parameter of the system of two identical fragment nuclei formed at the fission of nucleus with $A$ nucleons. The asymptotic value of the energy level density parameter of nucleus with $A$ nucleons has the volume and surface contributions related to coefficients $\alpha$ and $\beta$, respectively, and is written as \cite{ripl3}
\begin{eqnarray}
a(A) = \alpha A + \beta A^{2/3}.
\end{eqnarray}
The values of these coefficients obtained in the framework the back-shifted Fermi gas model are $\alpha=0.0722396$ MeV$^{-1}$ and $\beta=0.195267$ MeV$^{-1}\;$ \cite{ripl3}.

At $\beta=0$ Eqs. (2) -- (3) are simplified and transformed into
\begin{eqnarray}
a_{\rm sp}(A) = \alpha A , \\ \kappa = 0.
\end{eqnarray}
As a result, the expression for the width in this case is written as
\begin{eqnarray}
\sigma^2_{\rm S} = \frac{2 A^2 T}{C}.
\end{eqnarray}
This expression was obtained by Strutinsky \cite{s1964}. (Note that expression for width obtained in Ref. \cite{s1964} has another value of numerical coefficient, because another definition of the asymmetry of the fission fragments is used.)

The fission-fragment width $\sigma^2 \propto \frac{2 A^2 T}{C}=\sigma^2_{\rm S}$ increases linearly with temperature in the low temperature limit $2 \kappa A^{2/3} T^2 \ll C$. The width $\sigma^2$ has the maximum value at $C = 2 \kappa A^{2/3} T^2$. In the limit of extremely high temperatures $2\kappa A^{2/3} T^2 \gg C$ the width $\sigma^2 \sim \frac{A^{4/3} }{\kappa T}$ is inversely proportional to temperature. The influence of the surface energy level density parameter on the width of the fission-fragment mass yield rises with the number of nucleons in fissioning nucleus and temperature. In contrast to the width $\sigma^2$, the width $\sigma^2_{\rm S}$ increases linearly with temperature at any value of the temperature.

\section{Discussion of results and conclusions}

The bremsstrahlung spectrum is continuous; therefore, the nucleus can be excited by $\gamma$-quantum with the different energies in the experiment with the bremsstrahlung. However, the $\gamma$-fission cross-section is strongly rising with energy \cite{nr}. So, we consider that the excitation energy of fissioning compound nuclei $E^\star_{\rm cn}$ is very close to the end-point energy of the bremsstrahlung.

The excitation energy of fissioning compound nucleus $E^\star_{\rm cn}$ and the excitation energy of the fissioning system at the saddle point $E^\star_{\rm sp}$ related to the formation of the mass yield are connected by equation \cite{dms,dbkss}
\begin{eqnarray}
E^\star_{\rm cn} = E^\star_{\rm sp} + V_{\rm sp} - Q_{\rm fiss} = E^\star_{\rm sp} [1 + (V_{\rm sp} - Q_{\rm fiss})/E^\star_{\rm sp}].
\end{eqnarray}
Here $Q_{\rm fiss}$ is the $Q$-value of fission reaction on two symmetric fragments, and $V_{\rm sp}$ is the height of saddle point responsible for the yield of symmetric fission fragments. The values of $V_{\rm sp}$ and $Q_{\rm fiss}$ are close to each other \cite{dms}. As the result, $(V_{\rm sp} - Q_{\rm fiss})/ E^\star_{\rm sp} \ll 1$ for highly-excited nucleus, therefore $E^\star_{\rm cn} \approx E^\star$. Due to this the temperature of fissioning compound nucleus \begin{eqnarray}
T_{\rm cn} = \sqrt{E^\star_{\rm cn}/a(A)} = \sqrt{E^\star_{\rm cn}/(\alpha A + \beta A^{2/3})}
\end{eqnarray}
is close to the temperature of two-fragment system in the saddle point
\begin{eqnarray}
T = \sqrt{E^\star_{\rm sp}/a_{\rm sp}(A)}= \sqrt{E^\star_{\rm sp}/(\alpha A + 2^{1/3} \beta A^{2/3})}.
\end{eqnarray}
The ratio of these temperatures at the same excitation energy for $A \gtrsim 200$ belongs to the range
\begin{eqnarray}
0.96 \leq \frac{T}{T_{\rm cn}}=\sqrt{\frac{\alpha + \beta A^{-1/3}}{\alpha + 2^{1/3}\beta A^{-1/3}}} \leq 1.
\end{eqnarray}
The differences between both the excitation energy values $E^\star_{\rm cn}$ and $E^\star_{\rm sp}$ and the temperatures $T_{\rm cn}$ and $T$ are negligible. Nevertheless, small differences between these excitation energies and temperatures should be considered at careful consideration.

The evaporation of neutrons from the fissioning nuclei with $A\lesssim220$ is a very important process at high excitation energies \cite{o,expAu,expBi1,expBi2,expBi3,expBi4,expBi5,nnn}, because the values of the fission barrier in these nuclei are larger than the values of the neutron separation energy as a rule. Due to evaporation of many neutrons the most probable mass of the experimental fragment mass distribution $A_{\rm prob}$ is strongly smaller than $A/2$ at excitation energies higher than the fission barrier \cite{expAu,expBi1,expBi2,expBi3,expBi4,expBi5,o,nnn}. For an example, the value $A_{\rm prob}$ for the photo-fission of $^{197}$Au by the bremsstrahlung of end-point energy 1100 MeV is 92 \cite{expAu}, which is strongly smaller $A/2=197/2=98.5$. This means that the fission occurs after evaporation of around 13 neutrons as a rule. Note this number of evaporated neutrons is statistical averaging value. Due to neutron evaporation the number of nucleons in the fissioning system at the saddle point is close to $2A_{\rm prob}$. Therefore, we should substitute $A=2A_{\rm prob}$ into Eqs. (1) and (7). Note the experimental values of $A_{\rm prob}$ depend on the excitation energy and are extracted from an analysis of the the experimental fragment mass distributions \cite{expAu,expBi1,expBi2,expBi3,expBi4,expBi5,o,nnn}.

The excitation energy of the fissioning system is reduced due to the emission of pre-fission neutrons. Therefore, the average excitation energy of the nucleus at the moment of the scission can be determined as
\begin{eqnarray}
E^\star_{\rm eff} \approx E^\star_{\rm sp}-(A - 2 A_{\rm prob}) \tilde{E}_{\rm n}.
\end{eqnarray}
Here $\tilde{E}_{\rm n}$ is the average energy removed by the neutron from the fissioning nucleus at the evaporation of $A - 2 A_{\rm prob}$ neutrons. The value of $\tilde{E}_{\rm n}$ has two contributions related to the neutron binding energy and the average kinetic energy of the evaporated neutron
\begin{eqnarray}
E_{\rm kin} \approx \frac{3}{2}T_{\rm eff} \approx \frac{3}{2} (E^\star_{\rm eff}/a(2A_{\rm prob}))^{1/2}.
\end{eqnarray}
As a result, $\tilde{E}_{\rm n}$ can be approximated as
\begin{eqnarray}
\tilde{E}_{\rm n} \approx \frac{BE(A,Z)-BE(2A_{\rm prob},Z)}{A - 2 A_{\rm prob}} + \frac{3}{2}T_{\rm eff},
\end{eqnarray}
where $BE(A,Z)$ is the binding energy of nucleus with $A$ nucleons and $Z$ protons \cite{audi}.
Using Eqs. (12)-(14) we find the effective temperature
\begin{eqnarray}
T_{\rm eff}=\left[ \frac{E^\star_{\rm sp}-(BE(A,Z)-BE(2A_{\rm prob},Z))}{a(2 A_{\rm prob})} \right. \nonumber \\ \left. + \left( \frac{3(A - 2 A_{\rm prob})}{4 a(2 A_{\rm prob})} \right)^2  \right]^{1/2}- \frac{3(A - 2 A_{\rm prob})}{4 a(2 A_{\rm prob})}.
\end{eqnarray}
Note, the effective temperature should be used in Eqs. (1) and (7).

Note the values of $2A_{\rm prob}$ and $T_{\rm eff}$ can be evaluated in the framework complex statistical codes, which take into account the competition between the emission of neutrons and the fission. However, the substitutions $A$ on $2A_{\rm prob}$ and $T$ on $T_{\rm eff}$ in Eqs. (1) or (7) are very useful and strongly simplify the application of these equations. Moreover, we can compare the values of stiffness parameter $C$ obtained for various reactions. The effect of neutron emission on the values of $A$ and $E$ is not taken into account in Ref. \cite{dbkss}.

\begin{center}
\includegraphics[width=8.5cm]{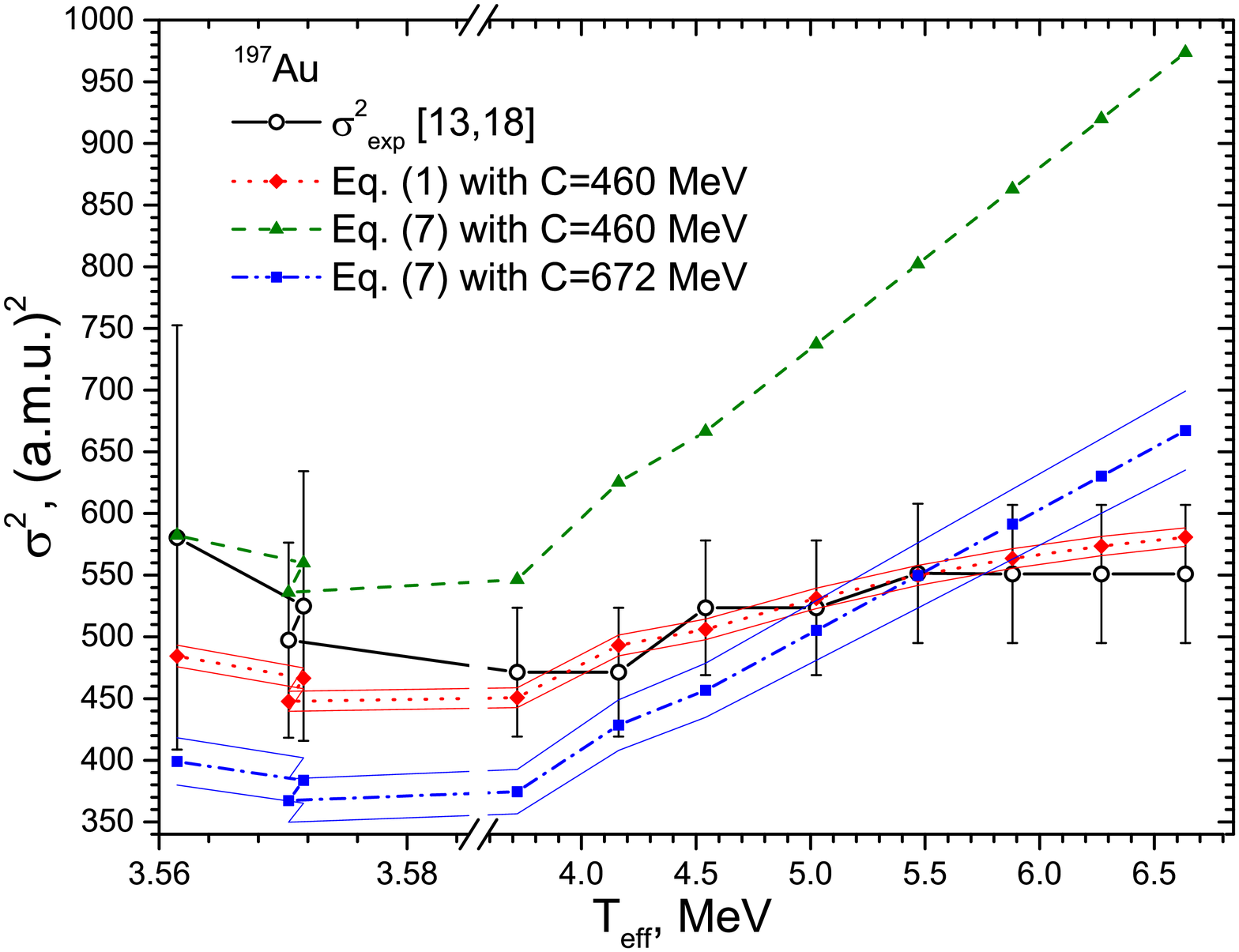}
\figcaption{\label{fig2}   The temperature dependence of the fission-fragment width $\sigma^2$ for nuclei $^{197}$Au evaluated by using Eqs. (1) and (7) at various values of the stiffness $C$. The thin lines show the range of uncertainty of the width induced by uncertainty of $C$. The experimental data are taken from Ref. \cite{expAu,expBi5}.}
\end{center}

The values of the width of fission-fragment mass yield evaluated using Eqs. (1) and (7) in our approach are compared with experimental data for $^{197}$Au and $^{209}$Bi in Figs. 1 and 2, respectively. The experimental data for the width and $A_{\rm prob}$ for $^{197}$Au obtained at the bremsstrahlung of end-point energies 300, 350, 400, 450, 500, 600, 700, 800, 900, 1000, and 1100 MeV are taken from Refs. \cite{expAu,expBi5}. The corresponding experimental data for $^{209}$Bi evaluated at the bremsstrahlung of end-point energies 40, 65, 85, 600, and 700 MeV are taken from Refs. \cite{expBi1,expBi2,expBi3,expBi4,expBi5} and the data at the bremsstrahlung of end-point energies 450, 500, 600, 700, 800, 900, 1000, and 1100 MeV are picked-up from Ref. \cite{expBi5}. (Note that the experimental value of the width of fission-fragment mass yield was also measured in the photo-fission of $^{209}$Bi target by the bremsstrahlung of end-point energies 2500 MeV in \cite{expBi6}. The excitation energy per nucleon at absorption of such high-energy gamma-quantum is larger than the binding energy per nucleon in $^{209}$Bi. So, the formation of fission fragments in such case should be affected by the various pre-equilibrium and non-equilibrium effects, therefore we ignore this data point in our analysis.)

\begin{center}
\includegraphics[width=8.5cm]{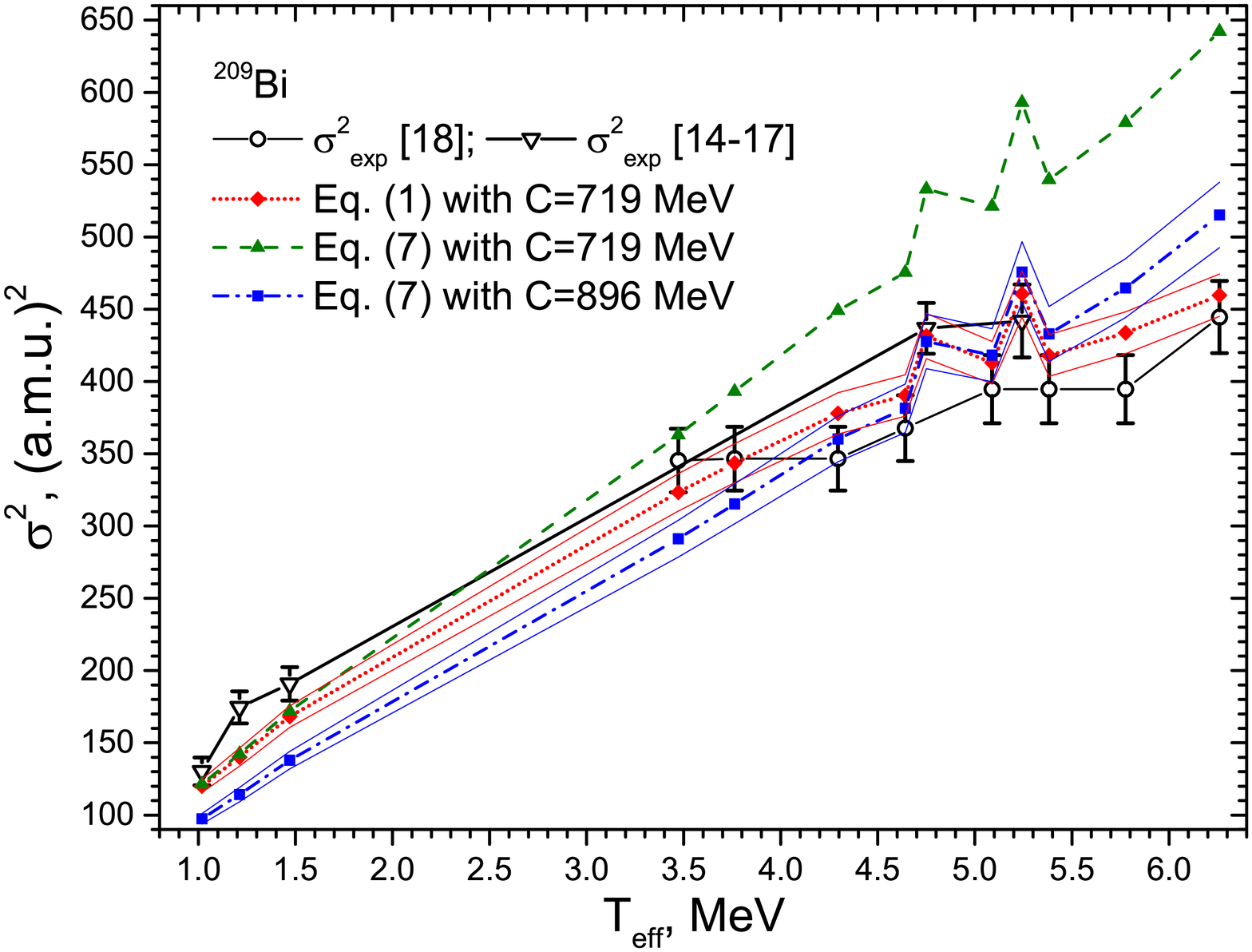}
\figcaption{\label{fig2}    The temperature dependence of the fission-fragment width $\sigma^2$ for nuclei $^{209}$Bi evaluated by using Eqs. (1) and (7) at various values of the stiffness $C$. The thin lines show the range of uncertainty of the width induced by uncertainty of $C$. The experimental data are taken from Refs. \cite{expBi1,expBi2,expBi3,expBi4,expBi5}.}
\end{center}

The experimental data-points from different Refs. and theoretical results are connected by lines. The dots are ordered according to the rising of the bremsstrahlung of end-point energy. As a rule, the temperature and the width of fission-fragment mass yield increase with the bremsstrahlung of end-point energy. Note the most probable fragment masses $A_{\rm prob}$ at fission of $^{197}$Au at the bremsstrahlung of end-point energies 300, 350, and 400 MeV are 97, 95, and 93 \cite{expAu,expBi5}. The corresponding effective temperatures $T_{\rm eff}$ at the saddle point are 3.561, 3.572, and 3.570 MeV. Due to such dependencies of $A_{\rm prob}$ and $T_{\rm eff}$ in the range 300$\div$400 MeV the width of fission-fragment mass yield decreases with the rising of the bremsstrahlung of end-point energy \cite{expAu,expBi5}. Such dependence of the width is out of the common trend, therefore we propose additional experimental studies of fission of $^{197}$Au at the bremsstrahlung of end-point energies 300, 350, and 400 MeV.

The values of the stiffness parameter of the potential related to the mass-asymmetric degree of freedom at the saddle point $C=460\pm 10$ MeV for $^{197}$Au and for $C=719\pm20$ MeV for $^{209}$Bi are found by the fitting the experimental data by Eq. (1). The uncertainties in $C$ are evaluated using the experimental uncertainties of values $\sigma$ and $A_{\rm prob}$ presented in Refs. \cite{expAu,expBi1,expBi2,expBi3,expBi4,expBi5}. The uncertainties in $C$ lead to the range of values of the fitted fission-fragment widths pointed by thin lines in Figs. 1 and 2.  The value of $C$ for $^{197}$Au is close to $C=382.5$ MeV obtained for slightly heavier nucleus $^{201}$Tl  at temperatures in the range $0.9$ MeV $ \lesssim T \lesssim 1.4$ MeV in Ref. \cite{dbkss}. However, the value of $C$ for $^{209}$Bi obtained now in an analysis of photo-fission is higher than the one evaluated in an analysis of the particle-induced fission for nearest nuclei $^{209}$Bi and $^{210}$Po in the range $0.8$ MeV $ \lesssim T \lesssim 1.4$ MeV in Ref. \cite{dbkss}. Note that wider temperature interval and experimental data for $2A_{\rm prob}$ used now lead to more accurate determination of the value of stiffness.

The energy dependence of the width of fission-fragment mass yield evaluated using Eq. (1) agree well with experimental data, see Figs. 1 and 2. If we substitute the found values of the stiffness parameters into Eq. (7) then the calculated values of the widths strongly overestimate the data at high energies. We remind that Eqs. (1) and (7) at the same value of the stiffness parameter $C$ lead to very close values of the width at small temperatures, see, for example, Fig. 2.

The quality of the description of the width can be estimated by using the deviation
\begin{eqnarray}
S =\frac{1}{N-1} \sum_{i=1}^N \left(\frac{\sigma^2_{i\;{\rm theor}} -\sigma^2_{i\;{\rm exp}}}{\Delta \sigma^2_{i\;{\rm exp}}} \right)^2 ,
\end{eqnarray}
where $\sigma^2_{i\;{\rm theor}}$ and $\sigma^2_{i\;{\rm exp}}$ are the theoretical and experimental values of the width in the point $i$, while $\Delta \sigma^2_{i\;{\rm exp}}$ is the error of the experimental value $\sigma^2_{i\;{\rm exp}}$. The values of $S$ evaluated by using Eqs. (1) and (7) for $^{197}$Au are 0.194 and 18.4, respectively. The corresponding values for $^{209}$Bi are 1.99 and 26.4.

We also find the values of $C$ by fitting the experimental data with the help of Eq. (7). The values of $C$ obtained in this case for $^{197}$Au and $^{209}$Bi are $672\pm32$ MeV and $896\pm41$ MeV, respectively. These values of $C$ are larger than the ones obtained previously. The uncertainties in $C$ leading to the uncertainties in the values of the corresponding widths are shown in Figs. 1 and 2. The values of $S$ evaluated for $^{197}$Au and $^{209}$Bi are 1.80 and 7.76, respectively. These values of $S$ are strongly larger than the ones evaluated using Eq. (1).

Comparing the values of $S$ obtained in various approaches and the results presented in Figs. 1 and 2 we conclude that Eq. (1) for the width describes the data in a wide range of the temperatures of the fissioning nucleus. In contrast to this, Eq. (7) for the width cannot describe the data in a wide range of the temperatures. As we have pointed, the uncertainties in $C$ lead to the uncertainties in the values of the corresponding widths. Taking into account the uncertainties in the calculated values of the widths marked in Figs. 1 and 2 we can conclude that in the framework of proposed approach the application of Eq. (1) leads to a better description of the widths than the using of Eq. (7).

In conclusion, Eq. (1) for the temperature dependence of the width of fission-fragment mass yield describes well the data for the bremstralung fission of $^{197}$Au and $^{209}$Bi at intermediate energies. The difference between our and the Strutinsky results takes place at high temperatures. The width of fission-fragment mass yield evaluated by using the volume and surface terms describes better data for high excitation energies than the one obtained at the same values of parameters and using the volume term only. The substitutions $A$ on $2A_{\rm prob}$ and $T$ on $T_{\rm eff}$ in Eq. (1) are very useful for an application of this equation in the case of the emission of pre-fission neutrons.

\end{multicols}

\clearpage

\end{CJK*}
\end{document}